# Synthesis and luminescence properties of electrodeposited ZnO Films

**C. V. Manzano[1], D. Alegre[1], O. Caballero-Calero[1], B. Alén[1], M. S. Martín-González[1]**

1. *IMM-Instituto de Microelectrónica de Madrid (CNM-CSIC), Isaac Newton 8, PTM, E-28760 Tres Cantos, Madrid, Spain*

*Corresponding author: marisol@imm.cnm.csic.es*

## ABSTRACT

ZnO films have been grown on gold (111) by electrodeposition using two different $OH^-$ sources, nitrate and peroxide, in order to obtain a comparative study between these films. The morphology, structural and optical characterization of the films were investigated depending on the solution used (nitrate and peroxide) and the applied potential. Scanning Electron Microscopy pictures show different morphologies in each case. X-Ray Diffraction confirms that the films are pure ZnO oriented along the (0002) direction. ZnO films have been studied by photoluminescence to identify the emission of defects in the visible range. A consistent model that explains the emissions for the different electrodeposited ZnO films is proposed. We have associated the *green and yellow emissions* to a transition from the donor $OH^-$ to the acceptor zinc vacancies ($V_{Zn}^-$) and to interstitial oxygen ($Oi^0$), respectively. The *orange-red emission* is probably due to transitions from the conducting band to $O_i^-$ and $O_{Zn}^0$ defects and the *infrared emission* to transition from these $O_i^{-/2-}$ and $O_{Zn}^{0/-}$ defects to the valence band.

# 1. INTRODUCTION

There has been a great interest in Zinc Oxide (ZnO) in recent years because it is a material with remarkable and varied properties. Due to its wide band gap (3.36 eV), it is transparent in the visible range. It has a large exciton energy (60 meV) which implies that excitonic laser action of ZnO can be observed above room temperature. In addition, it has a multicolored visible emission depending on the growth conditions[1]. Those characteristics provide an efficient emission in the ultraviolet and visible ranges, even at room temperature. Among other interesting properties, ZnO is: a piezoelectric material, has a great magneto-optical effect, behaves as a good chemical sensor[2,3], and is biocompatible and bio-safe[4,5]. Moreover, when doped with Al, it presents a large power factor (high figure of Merit ZT) at high temperatures when compared to other metal oxides[6]. More recently, this semiconductor has been the subject of intense research due to the appearance of ferromagnetism when it is "doped" with transition metals. These metals have been deemed room temperature diluted magnetic semiconductors, although it has been demonstrated that this effect could be also due to interfacial electrochemical reactions and interfaces effects[7-12]. As a consequence of those varied properties, ZnO can be used in: cantilevers for Atomic Force Microscopes (AFM)[13], dye sensitized solar cells[14,15], sun creams and as a white light emitting diodes (LEDs)[16,17]. Finally, several groups have fabricated homojunctions with ZnO, obtaining emission of different colors such as blue-violet[18] and blue-yellow[19]. Also, UV photodiodes have been reported in the literature[20]. Other colors observed from heterojunctions with ZnO were UV[21,22], violet-white[23], blue-white and and white[24].

As far as the growth method is concerned, there are multiple ways of obtaining ZnO films, such as Chemical Vapour Deposition (CVD), and Metal-Organic CVD (MOCVD), Sputtering, Molecular Beam Epitaxy (MBE), and Pulsed Laser Deposition (PLD). The main drawback of the above mentioned techniques is that they require ultra-high vacuum and in some cases, high temperatures. A way to avoid these constraints is to use other kind of fabrication methods, such as Chemical Bath Deposition (CBD) or electrodeposition. In this sense, doped and un-doped ZnO have been grown by electrodeposition in a successful and controlled way[25]. The electrodeposition is a large area approach, which provides, in principle, the same type of emitting ZnO nanorods as the aqueous chemical growth since it is also a low temperature process. However, it has the advantage of selective area growth the by pre-structuring of the metallic substrate; therefore it is of great interest for applications requiring the integration of LEDs in specific areas. This factor reduces one of the processing steps when making ZnO nanorod based devices.[12]

Electrodeposition of ZnO is usually performed using solutions with an acidic pH. The process implies the electrochemical formation of hydroxide ions at the surface of the working electrode. These ions react with the zinc ions present in the solution to form zinc oxide by means of the dehydration of zinc hydroxide at temperatures greater than 50ºC[26]. The general reaction is:

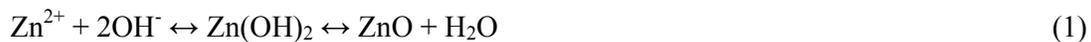

$$Zn^{2+} + 2OH^- \leftrightarrow Zn(OH)_2 \leftrightarrow ZnO + H_2O \qquad (1)$$

The hydroxide ions ($OH^-$), which act as the precursors for the formation of zinc hydroxide, can be electrochemically obtained via three different ways[26,27]:

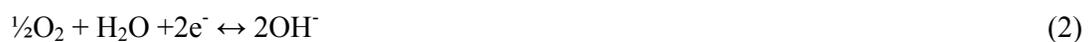

$$\tfrac{1}{2}O_2 + H_2O + 2e^- \leftrightarrow 2OH^- \qquad (2)$$

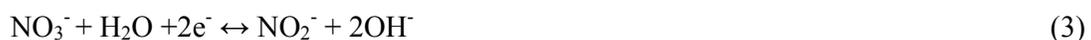

$$NO_3^- + H_2O + 2e^- \leftrightarrow NO_2^- + 2OH^- \qquad (3)$$

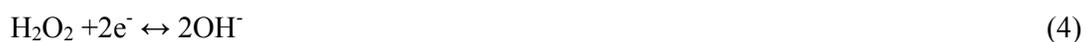

$$H_2O_2 + 2e^- \leftrightarrow 2OH^- \qquad (4)$$

The first studies made about the electrochemical deposition of zinc oxide were published at the same time by two independent groups, (Lincot et al.[28]) who used dissolved oxygen and (Izaki et al.[29]) who used nitrate. Later on, Lincot et al. developed and studied the mechanism of another different precursor: hydrogen peroxide[27,30]. This precursor presents two main advantages over the ones previously used: larger deposition velocity due to its great solubility in water (which was the main disadvantage when dissolved oxygen was used) and it does not generate any secondary reaction product (which was a problem when nitrate ions were present). Nevertheless, all these reactions are reversible, giving rise to quasi-equilibrium conditions along the growth process. This explains the high crystalline quality achieved by electrodeposition of ZnO, quite similar to that obtained by means of other more expensive techniques, as the ones previously mentioned. Also, it is interesting to note that the production of the oxide is direct without further thermal treatment as it is normal for most oxides[31,32].

In this work, we present, to the best of our knowledge, the first comparative study of the quality of the films using different $OH^-$ sources. Both the morphology and the photoluminescence emission of electrodeposited ZnO films have been studied. Furthermore, a consistent model that explains the luminescence emissions of the defects present in electrodeposited ZnO films is also presented.

## 2. EXPERIMENTAL

The electrochemical deposition has been performed with a standard electrochemical cell with a three electrode configuration: a platinum mesh as anode, a silver-silver chloride (Ag/AgCl 3M) electrode as reference electrode ($E_0 = + 0.208$ V versus Normal Hydrogen Electrode, NHE), and a thin film of 150 nm gold (111) electron-beam evaporated on a silicon substrate (100) working electrode or cathode. The cell was controlled with a bi-potentiostat (Eco Chemie, Model AUT302.0) computer operated.

In order to perform the electrodeposition of the ZnO thin films, two different solutions of high purity (99.999%, Sigma-Aldrich Co.) zinc salts have been used. The other reactants are reagent grade potassium chloride from Panreac Química, S.A.U. and reagent grade stabilized 30 % hydrogen peroxide from Panreac Química, S.A.U. The *nitrate solution contained* 0.1 M $Zn(NO_3)_2$ in deionized water (< 14 mS) and the p*eroxide solution was prepared by* 5mM $ZnCl_2$ + 0.04 M $H_2O_2$ + 0.1M KCl in deionized water (< 14 mS).

The ZnO films were grown in a thermostatic bath at 80ºC with a thermal stability of ± 1 ºC. The electrodeposition was carried out in the following conditions:

a)  Nitrate solution: ZnO films were deposited potentiostatically at -1.0, -0.85 V, -0.7 V and -0.6 V vs. Ag/AgCl for one hour.

b)  Peroxide solution: ZnO films were deposited potentiostatically for one hour at different potentials. The chosen potentials were -0.9 V, -0.7 V, -0.5 V and -0.3V.

With regard to the characterization techniques used in this work, they can be divided into two groups. Firstly, the morphology characterization of the films was performed using Scanning Electron Microscopy (SEM) with a Hitachi S-800 and X-Ray Diffraction (XRD) Spectroscopy with a Philips X-Pert Cuα X-Ray transmitter four circle diffractometer. Secondly, the defect characterization was carried out via photoluminescence (PL) measurements. The samples were excited with a pulsed Nd:YAG tripled laser (355 nm wavelength, 15 ns pulse, 20 kHz repetition rate) modulated in intensity with an optical modulator (177 Hz). Then, the light emitted in the ultraviolet and/or visible range was filtered by suitable long pass filters and dispersed by a monochromator with 300 mm focal length (diffraction grating: 1200 lines/mm) and detected by means of a cooled photomultiplier, connected to a lock-in amplifier.

# 3. RESULTS AND DISCUSSION.

## 3.1 ZnO Deposition

As mentioned before, two different sources have been employed to obtain the hydroxide ions ($OH^-$) at the surface of the cathode: $NO_3^-$ and $H_2O_2$. Both cyclic voltammograms are shown on Fig. 1.

In the case of the nitrate solution, Fig 1.(a), the deposition process of zinc hydroxide from a zinc nitrate solution is based on the reduction of the nitrate. A reduction peak at potentials around -0.85 V vs. Ag/AgCl is observed. This peak should be related to the reduction of $NO_3^-$.

The thermodynamically stable specie of this solution is ammonium ion ($NH_4^+$)[33] at a pH of 4.5. However, the presence of $NO_2^-$ is the specie normally described in the literature,[34-39] according to the reaction:

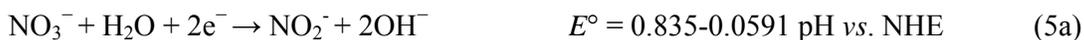
$$NO_3^- + H_2O + 2e^- \rightarrow NO_2^- + 2OH^- \qquad E° = 0.835\text{-}0.0591 \text{ pH } vs. \text{ NHE} \qquad (5a)$$

Although $NO_2^-$ is not thermodynamically stable under the present conditions, the kinetic of further reduction to ammonium ion is slow. However, the reaction (5b) can occur with time producing more $OH^-$ according to the reaction with time:

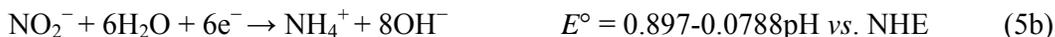
$$NO_2^- + 6H_2O + 6e^- \rightarrow NH_4^+ + 8OH^- \qquad E° = 0.897\text{-}0.0788\text{pH } vs. \text{ NHE} \qquad (5b)$$

Thus, the mechanism for cathodic reduction of nitrate follows a consecutive reaction pathway $NO_3^- \rightarrow NO_2^- \rightarrow NH_4^+$. Therefore, reduction to ammonium ions should be also taken into account for long deposition times. In fact, Yu et al.[40] studied the reduction of $NO_3^-$ ions at room temperature and concluded that nitrite ($NO_2^-$) accumulation was favored at less negative potentials and shorter deposition times, while ammonium formation was favored at more negative potentials and longer deposition times. The kinetics of disappearance of nitrite in our case should be faster than in that work since in our case the experiment is performed at 80ºC to obtain the ZnO directly.

After the reduction peak at -0.85 V, the diminishing of the current in the voltammogram is attributed to the precipitation of ZnO observed experimentally on the electrode surface. In this case, different films were deposited at -1.0, -0.85 V, -0.7V and -0.6 V vs. Ag/AgCl at 80ºC and in all of them the presence of pure ZnO oriented along [0001] direction was detected by XRD (see Fig. 3 and the discussion of next section). No deposit was obtained at potentials less negative than -0.5 V. Only the more representative results are presented in this study.

In the case of the peroxide solution, Fig 1.(b), the deposition process of zinc hydroxide is based on the reduction of the $H_2O_2$. This reduction takes place initially at pH = 5.8. A possible general reduction could be:

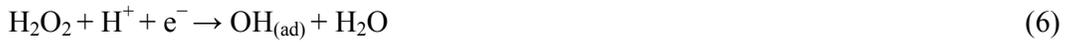

$$H_2O_2 + H^+ + e^- \rightarrow OH_{(ad)} + H_2O \qquad (6)$$

where (ad) stands for adsorbed. This Eq. is reported in the literature as a step process followed by a final desorption step[27]:

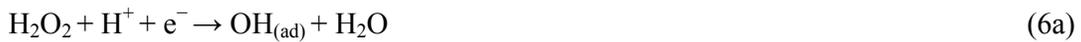

$$H_2O_2 + H^+ + e^- \rightarrow OH_{(ad)} + H_2O \qquad (6a)$$

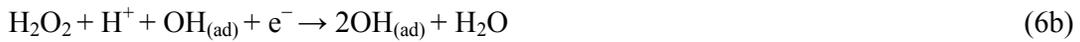

$$H_2O_2 + H^+ + OH_{(ad)} + e^- \rightarrow 2OH_{(ad)} + H_2O \qquad (6b)$$

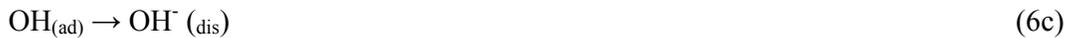

$$OH_{(ad)} \rightarrow OH^-_{(dis)} \qquad (6c)$$

where (dis) stands for in solution.

In this case different films were deposited at -0.9, -0.7 V, -0.5 V and -0.3V vs. Ag/AgCl at 80ºC and in them the presence of pure ZnO were detected by XRD. No deposit was obtained at less negative potentials than -0.2 V. Only the more representative results are presented in this study.

### 3.2 Quality and properties of ZnO films obtained.

#### 3.2.1 Nitrate solution

SEM micrographs of the films electrodeposited with the nitrate solution are shown in Fig. 2. As it can be seen there, the morphology of the deposited film depends on the applied potential during deposition. For the most negative potentials (-1 V, Fig. 2(a)) rough films are obtained. That means that the growth was quick and it had formed disordered columns as show in the cross section (Fig. 2 (b)). On the other hand, when the applied potential is less negative (-0.7 V, Fig. 2(c and d), the obtained film is thinner, more compact and presents a clearer order. The morphology of the film at -0.7 V presents columnar and two-dimensional growth. For the least negative potential in which it has been possible to obtain a deposit (-0.6 V, Fig. 2(e and f)), the surface of the film is the smoothest. As can be seen in Fig. 2 (a, c and e) the thickness of the deposit was around 30 microns for -1 V, around 4 microns for -0.7 V and around 0.4 microns for -0.6V.

Fig. 3 shows the X-Ray Diffraction patterns of the ZnO films deposited in the above mentioned conditions. The intensity has been displayed in log scale to enhance small diffractions peaks. It can be seen that the film is purely made of ZnO, because the peaks to be seen are only the ones from the substrate (Si with a 150 nm Au layer) and two of ZnO, (0002) and (0004), indicating that only one crystalline direction of ZnO is present in the film. The films are textured along the [001] direction, due to the coalescence of individual hexagonal grains. The full width half maximum (FWHM) of these curves are presented in Table 1. The FWHM is smaller (0.115 vs. 0.17) for films electrodeposited at the more negative potential. This is related to the fact that more negative potentials give bigger ZnO grain sizes.

*3.2.2 Peroxide solution*

The morphology of the films obtained with this solution can be seen in Scanning Electron Microscopy images (Fig. 4) for different potentials. The film grew at the most negative potential (- 0.9 V, Fig. 4 (a)) presents hexagonal columns. These columns are aligned perpendicularly to the substrate´s surface, as it is shown in the cross sectional micrograph of this film in Fig. 4(b), where a clear columnar growth can be seen. This is due again to the coalescence of individual hexagonal grains. In other words, it is a nucleation-growth mechanism. Fig. 4 (c and d) displays the SEM images of the ZnO film electrodeposited at -0.5 V. The cross section (Fig. 4 (d)) shows a columnar growth, but the grains are not hexagonal as in the case of the nitrite solution (see top view Fig . 4(c)). For the least negative potential (-0.3V, Fig. 4 (e and f)), the film obtained has the smoothest surface found for this solution, and columns are not observed. For lower potentials than -0.3V no deposit was obtained.

In order to study the crystalline quality of the films grown in the peroxide solution, XRD were taken (see Fig. 5). As in the previous case, only the diffraction peaks of ZnO in the [0001] direction can be seen for the films electrodeposited at -0.9 V, -0.5 V and -0.3V. That means that the films are textured along this direction. However, the XRD pattern of the film electrodeposited at -0.3 V shows only one peak of ZnO, (0002). The (0004) peak is not observed in the XRD pattern.  Table 1 lists the values of the FWHM of ZnO films electrodeposited in peroxide solution.  As in the previous case, the film electrodeposited at the least negative potential (-0.3 V) presents smaller grain size (bigger FWHM).

It is interesting to note here that the films obtained at more negative potentials is thinner for the same total area than films grown at less negative potentials (see Fig. 4 (b), (d) and (f)).

A possible explanation for that is in the case of electrochemical deposition performed at very negative potentials and/or large hydrogen peroxide concentration, then Eq. (6c) will be the dominant one. In these cases, the $OH^-$ formed at the surface of the electrode is desorbed from the electrode and will react in the solution with $Zn^{2+}$ forming $Zn(OH)_2$ powder out of the electrode, which will react at temperature > 50ºC to form ZnO. The experimental observation of this reaction is obtained by looking at the electrodeposition cell during deposition. A white precipitate appears in the solution going from a position close to the working electrode to the bottom of the electrochemical cell. This causes a very slow but rather ordered growth. The white precipitate does not appear at less negative potentials. As it can be seen in Fig. 4 (b) and 4 (d) the thickness of the deposit for the same area (0.196 $cm^2$) was around 0.3 microns for -0.9 V and around 0.6 microns for -0.3V. This highlights that the Faradaic efficiency of the system is higher for -0.3 V, since for more negative potentials not all the $OH^-$ produced at the electrode surface generates ZnO deposit.

### 3.3 Crystalline defects: photoluminescence.

In the case of ZnO, the band-gap emission appears in the near ultraviolet around 370 – 385 nm (3.3 eV). All the studied films show photoluminescence (PL) in the visible region between 400 – 900 nm, as shown in Fig. 6. This broad band can be related to the presence of different defects (and/or impurities) with different concentrations in each sample. A deconvolution into four Gaussian peaks has been used in the bibliography[31] before to describe this visible emission in ZnO: a green component centered at 515 nm (2.41 eV); a yellow at 560 nm (2.21 eV); an orange-red centered at 670 nm (1.85 eV) and a tail in the infrared centred at 785 nm (1.58 eV). The resulting areas for each component are presented in Fig. 7 for the different samples of our study.

In the literature, there is a great controversy in the assignation of defects to the different emission lines (see Ref.[41] for instance). This disagreement is mainly due to two reasons: a) the wide and rather unstructured character of the emission between 400 – 900 nm; and b) many defects can be present in the sample at the same time[41,42].

There are many theoretical studies about the transition levels of the defects in $ZnO^{43-46}$. Among them, we have chosen the most recent and complete study presented by Van de Walle[42] in order to explain our results. According to this study, the typology of defects that most probably could be found in ZnO films grown under oxygen-rich conditions are intrinsic defects like zinc vacancies ($V_{Zn}$), interstitial oxygen ($O_i$), and oxygen anti-sites ($O_{Zn}$)[42]. Moreover, taking into account that the films grown by electrodeposition are generated according to reaction (1), extrinsic defects like trapped $OH^-$ should also be taken into account. This type of defect is known as $H-I^{47}$.

Fig. 8 represents the proposed model for the photoluminescence emissions of the electrodeposited ZnO films. It is important to note that the energy of the donor extrinsic defect $OH^-$ (H-I) has also been added to the graph from Ref. [46,47] for completeness. The proposed transitions have been marked as arrows and identified with their emission colour.

The *green and yellow emissions* are tentatively assigned to a transition from the donor $OH^-$ to the acceptor zinc vacancies ($V_{Zn}^-$) and interstitial oxygen ($O_i^0$), respectively. Since recent results suggest that both yellow and green emissions are present in samples grown in solution, they are both related to a transition from the same donor to different acceptors placed at different depths in the energy level scheme[48]. The existence of this kind of shallow donors has been long known in samples grown in solution[49], but its nature is more difficult to clarify. They have been related to $V_o^{49,50}$, but the theoretical calculations by Van de Walle[42] indicate that $V_o$ defects form rather deep levels, as it can be seen in Fig. 8. Given that no other intrinsic defect appears at a shallow position in the energy scheme, this characteristic must be attributed to an extrinsic defect. As it was mentioned before, the only extrinsic defects expected in our samples are hydrogen related defects, like the H-I. The energy proposed in the literature for this defect is 2.91 eV[46,47], which means that it can be considered as a shallow donor. Moreover, the best evidence that can be given is the absence of the yellow band and most of the green band, in the spectrum emitted from the sample that has been electrodeposited at the slowest growth rate (-0.9 V in the peroxide solution), as it is shown in Fig. 7. Due to the $OH^-$ desorption under these conditions (see Eq.6(c)), this sample is expected to have the smallest amount of trapped $OH^-$, and therefore of H-I defects. It should be noted that while the yellow emission disappears completely, the green band is not fully attenuated in the spectrum. Recent works have reported, via Positron Annihilation Spectroscopy (PAS)[51,52], the presence of zinc vacancies ($V_{Zn}^-$) surrounded by a variable number of OH bonds in these kind of samples. This justifies the choosing of $V_{Zn}^-$ as the acceptor in the green emission transition, because $OH^-$ stabilizes this kind of vacancies. Our association of the green and yellow emissions to transitions from the H-I defect (or trapped $OH^-$) to $Oi^0$ and $V_{Zn}^-$, respectively, can be further clarified by annealing the samples at 110ºC. At this temperature the H-I defects start to disappear slowly[48] with the subsequent reduction in the green and yellow emission as it is shown in Fig.9 and Fig. 10. Raising the annealing temperature to 150 ºC, the H-I defects are further reduced and at temperatures higher than 250ºC, the green emission is not longer observed.

The *orange-red emission* corresponding to an energy of 1.85 eV (670 nm) has been traditionally assigned to an excess of oxygen[53]. In our case, we have always attributed the excess of oxygen to the growth technique, but the literature does not specify which of the multiple oxygen defects present in ZnO is implied, only that the electron is de-excited from the conduction band[51]. Taking into account the theoretical transition values shown in Fig. 8[42], a transition of 1.85 eV is found for electrons coming from the conduction band towards $O_i^-$ and $O_{Zn}^0$ defects (the transition is represented with orange arrows in Fig. 8). In our samples, both defects are expected to be present and, given that their energy levels are quite similar, we can conclude that both of them can be responsible for this emission. The films electrodeposited at slower growth rates (potential of -0.9 V for the peroxide solution and -0.7 V for the nitrate solution) present a more important orange-red component (see Fig. 7). This could be explained if one or both $O_{Zn}$ and $O_i$ are favoured at slower deposition velocities.

The *infrared emission* corresponding to energy of 785 nm (1.58 eV) might be due to a transition from the defects $O_i^{-/2-}$ and $O_{Zn}^{0/-}$ to the valence band or a transition from the conduction band to $O_{Zn}^{-/2-}$, or even the transition from the H-I defect to $O_i^{-/2-}$ and $O_{Zn}^{0/-}$. All of them have an energy around the infrared values obtained, which are represented in Fig. 8 as red arrows. As it can be seen there, these infrared transitions complete the de-excitation path of the electrons that made a first de-excitation transition by emitting orange-red light. This infrared emission increases at annealing temperatures above 350ºC (see Fig. 10). This can be explained following the Eq.:

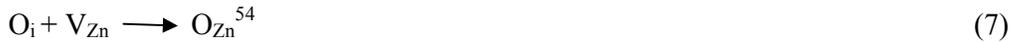

$$O_i + V_{Zn} \longrightarrow O_{Zn}^{54} \tag{7}$$

This new $O_{Zn}$ defect created as a consequence of the annealing is responsible for the emission increase found in our experiments.

*The band gap emission* was detected in the spectra of all samples with exactly the same energy (3.363 eV) (see Fig. 9). It is worth mentioning that this ultraviolet emission is quite rare in the case of electro-deposited samples. The explanation of its appearance at room temperature can be the presence of bound excitons. Its contribution increases when the samples are annealed. Moreover, when the annealing temperature is 250º C, the emission is enhanced 40 times. For higher temperatures the band gap emission starts to decrease probably due to the creation of new defects, possibly Vo[51].

In summary, we have associated the *green and yellow emissions* to a transition from the donor $OH^-$ to the acceptor zinc vacancies ($V_{Zn}$-) and interstitial oxygen ($Oi^0$), respectively. The *orange-red emission* is probably due to transitions from the conducting band to $O_i^-$ and $O_{Zn}^0$ defects and the *infrared emission* to transition from these $O_i^{-/2-}$ and $O_{Zn}^{0/-}$ defects to the valence band.

## 4. CONCLUSIONS.

In this paper we have obtained high quality films of ZnO growth via electrodeposition at constant potentials from two different solutions (nitrate and peroxide). The reactions that give rise to the electrodeposition process have been described. The morphological analysis of the grown films showed good control over their morphology depending on the electrodeposition conditions. X-Ray Diffraction spectra showed that the films were oriented with the c-axis perpendicular to the substrate (along the (0001) direction). Moreover, the defects present in the different films have been studied with photoluminescence experiments. The possible origins of the different emission transitions that are present in these ZnO electrodeposited films have been determined, with the aid of their relative intensities after the deconvolution of the complete spectra and their behaviour after annealing treatments. The defects present in our films are H-I defects, zinc vacancies and interstitial oxygen and anti-site oxygen. The first one acts as a shallow donor that gives rise to transitions that revert in green light emission when combined with zinc vacancies, and yellow light emission when interstitial oxygen act as acceptors. Then, the orange-red emission consists of transitions between the conduction band and interstitial or anti-site oxygen. Finally, the infrared emission consists in transition from interstitial oxygen or anti-site oxygen defects to the valence band. It is also worth mentioning that these films also present emission in the ultraviolet range, which is quite difficult to obtain in the case of electrodeposited samples. This transition can be associated with the presence of bound excitons in the as-grown films, where its contribution is greatly enhanced when annealed at 250ºC.

# 5. ACKNOWLEDGMENTS

Authors would like to acknowledge financial support form: MICINN project MAT2008-06330 and ERC StG NanoTEC 240497 and C.V.M. wants to acknowledge CSIC for JAE pre-doc, and O.C.C. wants to acknowledge CSIC for JAE-doc.

# 6. REFERENCES

[1] C. H. Ahn, Y. Y. Kim, D. C. Kim, S. K. Mohanta, and H. K. Cho, Journal of Applied Physics **105,** 013502 (2009).

[2] H. T. Wang, B. S. Kang, F. Ren, L. C. Tien, P. W. Sadik, D. P. Norton, S. J. Pearton, and J. Lin, Applied Physics Letters **86,** 243503 (2005).

[3] S. M. Al-Hilli, R. T. Al-Mofarji, P. Klason, M. Willander, N. Gutman, and A. Sa'ar, Journal of Applied Physics **103,** 014302 (2008).

[4] U. Ozgur, Y. I. Alivov, C. Liu, A. Teke, M. A. Reshchikov, S. Dogan, V. Avrutin, S.-J. Cho, and H. Morkoc, Journal of Applied Physics **98,** 041301 (2005).

[5] C. Klingshirn; *Vol. 244* (WILEY-VCH Verlag, 2007), p. 3027.

[6] T. Tsubota, M. Ohtaki, K. Eguchi, and H. Arai, Journal of Materials Chemistry **7,** 85 (1997).

[7] M. A. Garcia, A.Quesada, J.L. Costa-Krämer, J. F. Fernández, S. J. Khatib, A. Wennberg, A. C. Caballero, M. S. Martín-González, M. Villegas, F. Briones, J. M. González-Calbet, A. Hernando., Physical Review Letters **94,** 217206 (2005).

[8] A. Serrano, E. F. Pinel, A. Quesada, I. Lorite, M. Plaza, L. Pérez, F. Jiménez-Villacorta, J. de la Venta, M. S. Martín-González, J. L. Costa- Krämer, J. F. Fernandez, and J. G. Llopis, M. A., Physical Review B **79,** 144405 (2009).

[9] J. F. Fernández, A. C. Caballero, M. Villegas, S. J. Khatib, M. A. Bañares, J. L. G. Fierro, J. L. Costa-Kramer, E. Lopez-Ponce, M. S. Martín-González, F. Briones, A. Quesada, M. García, and A. Hernando, Journal of the European Ceramic Society **26,** 3017 (2006).

[10] M. S. Martin-Gonzalez, J. F. Fernandez, F. Rubio-Marcos, I. Lorite, J. L. Costa-Kramer, A. Quesada, M. A. Banares, and J. L. G. Fierro, Journal of Applied Physics **103,** 083905 (2008).

[11] M. S. Martin-Gonzalez, M. A. Garcia, I. Lorite, J. L. Costa-Kramer, F. Rubio-Marcos, N. Carmona, and J. F. Fernandez, Journal of the Electrochemical Society **157,** E31 (2010).

[12] M. S. Martín-Gonzalez, C. S. Steplecaru, F. Briones, E. López-Ponce, J. F. Fernández, M. A. García, A. Quesada, C. Ballesteros, and J. L. Costa-Krämer, Thin Solid Films **518**, 4607 (2010).

[13] W. Lee, M.-C. Jeong, and J.-M. Myoung, Acta Materialia **52**, 3949 (2004).

[14] S. Yanagida, G. K. R. Senadeera, K. Nakamura, T. Kitamura, and Y. Wada, Journal of Photochemistry and Photobiology A: Chemistry **166**, 75 (2004).

[15] Y. Gao and M. Nagai, Langmuir **22**, 3936 (2006).

[16] M. Willander and et al., Nanotechnology **20**, 332001 (2009).

[17] S. Kishwar, K. ul Hasan, N. H. Alvi, P. Klason, O. Nur, and M. Willander, Superlattices and Microstructures **49**, 32 (2011).

[18] Z. P. Wei, Y. M. Lu, D. Z. Shen, Z. Z. Zhang, B. Yao, B. H. Li, J. Y. Zhang, D. X. Zhao, X. W. Fan, and Z. K. Tang, Applied Physics Letters **90**, 042113 (2007).

[19] W. Liu, S. L. Gu, J. D. Ye, S. M. Zhu, S. M. Liu, X. Zhou, R. Zhang, Y. Shi, Y. D. Zheng, Y. Hang, and C. L. Zhang, Applied Physics Letters **88**, 092101 (2006).

[20] Y. R. Ryu, T. S. Lee, J. A. Lubguban, H. W. White, Y. S. Park, and C. J. Youn, Applied Physics Letters **87**, 153504 (2005).

[21] A. Tsukazaki, A. Ohtomo, T. Onuma, M. Ohtani, T. Makino, M. Sumiya, K. Ohtani, S. F. Chichibu, S. Fuke, Y. Segawa, H. Ohno, H. Koinuma, and M. Kawasaki, Nat Mater **4**, 42 (2005).

[22] Y. Ryu, T.-S. Lee, J. A. Lubguban, H. W. White, B.-J. Kim, Y.-S. Park, and C.-J. Youn, Applied Physics Letters **88**, 241108 (2006).

[23] T. Aoki, Y. Hatanaka, and D. C. Look, Applied Physics Letters **76**, 3257 (2000).

[24] Z. Z. Ye, J. G. Lu, Y. Z. Zhang, Y. J. Zeng, L. L. Chen, F. Zhuge, G. D. Yuan, H. P. He, L. P. Zhu, J. Y. Huang, and B. H. Zhao, Applied Physics Letters **91**, 113503 (2007).

[25] T. Pauporté and D. Lincot, Electrochim. Acta **45**, 3345 (2000).

[26] S. Peulon and D. Lincot, Journal of the Electrochemical Society **145**, 864 (1998).

[27] T. Pauporté and D. Lincot, Journal of Electroanalytical Chemistry **517**, 54 (2001).

[28] S. Peulon and D. Lincot, Advanced Materials **8**, 166 (1996).

[29] M. Izaki and T. Omi, Applied Physics Letters **68**, 2439 (1996).

[30] T .Pauporté and D. Lincot, J. Electrochem. Soc. **148**, C310 (2001).

[31] M. S. Martin-Gonzalez, J. Garcia-Jaca, E. Moran, and M. A. Alario-Franco, Journal of Materials Chemistry **9**, 137 (1999).

[32] M. Martin-Gonzalez, E. Moran, O. Rodriguez de la Fuente, and M. A. Alario-Franco, Journal of Materials Chemistry **11**, 616 (2001).

[33] M. J. N. Pourbaix, In Atlas of Electrochemical Equilibria in Aqueous Solutions, Pergamon, New York (1966).

[34] T. Mahalingam, V. S. John, M. Raja, Y. K. Su, and P. J. Sebastian, Solar Energy Materials and Solar Cells **88**, 227 (2005).

[35] M. Lai, Chem. Mater. **18**, 2233 (2006).

[36] S. Karuppuchamy, K. Nonomura, T. Yoshida, T. Sugiura, and H. Minoura, Solid State Ionics **151**, 19 (2002).

[37] X. Han, R. Liu, W. Chen, Z. Xu, Thin Solid Films **516** 4025 (2008).

[38] T. Ren, H. R. Baker, and K. M. Poduska, Thin Solid Films **515**, 7976 (2007).

[39] T. Yoshida, D. Komatsu, N. Shimokawa, and H. Minoura, Thin Solid Films **451-452**, 166 (2004).

[40] J. Yu and M. Kupferle, Water Air Soil Pollution: Focus **9**, 245 (2009).

[41] A. B. Djurišić and Y. H. Leung; *Vol. 2* (WILEY-VCH Verlag, 2006), p. 944.

[42] A. Janotti and C. G. Van de Walle, Physical Review B **76**, 165202 (2007).

[43] P. S. Xu, Y. M. Sun, C. S. Shi, F. Q. Xu, and H. B. Pan, Nuclear Instruments and Methods in Physics Research Section B: Beam Interactions with Materials and Atoms **199**, 286 (2003).

[44] S. A. M. Lima, F. A. Sigoli, M. Jafelicci Jr, and M. R. Davolos, International Journal of Inorganic Materials **3,** 749 (2001).

[45] B. Lin, Z. Fu, Y. Jia, and G. Liao, Journal of the Electrochemical Society **148,** G110 (2001).

[46] E. V. Lavrov, J. Weber, ouml, F. rrnert, C. G. Van de Walle, and R. Helbig, Physical Review B **66,** 165205 (2002).

[47] E. V. Lavrov, F. Börnert, and J. Weber, Physical Review B **72,** 085212 (2005).

[48] R. B. Lauer, Journal of Physics and Chemistry of Solids **34,** 249 (1973).

[49] J. F. C. J.C. Simpson, Journal Applied Physics **63** (1998).

[50] K. Kuriyama, M. Ooi, K. Matsumoto, and K. Kushida, Applied Physics Letters **89,** 242113 (2006).

[51] J. Cizek, N. Zaludova, M. Vlach, S. Danis, J. Kuriplach, I. Prochazka, G. Brauer, W. Anwand, D. Grambole, W. Skorupa, R. Gemma, R. Kirchheim, and A. Pundt, Journal of Applied Physics **103,** 053508 (2008).

[52] G. Brauer, J. Kuriplach, J. Cizek, W. Anwand, O. Melikhova, I. Prochazka, and W. Skorupa, Vacuum **81,** 1314 (2007).

[53] A. B. Djurisic, Y. H. Leung, K. H. Tam, L. Ding, W. K. Ge, H. Y. Chen, and S. Gwo, Applied Physics Letters **88,** 103107 (2006).

[54] Y. Yang, H. Yan, Z. Fu, B. Yang, L. Xia, Y. Xu, J. Zuo, and F. Li, Solid State Communications **138,** 521 (2006).

### Table Caption

*Table 1. Parameters from XRD analysis.*

| Solution | E (V) | FWHM |
|----------|-------|------|
| Nitrate  | -1.0  | 0.115 |
| Nitrate  | -0.7  | 0.158 |
| Nitrate  | -0.6  | 0.170 |
| Peroxide | -0.9  | 0.158 |
| Peroxide | -0.5  | 0.170 |
| Peroxide | -0.3  | 0.172 |

### Figure Caption

*Fig. 1. Linear voltammetric curves of (a) solution in 0.1M $Zn(NO_3)_2$ and (b) in 5mM $ZnCl_2 + 0.04M\ H_2O_2 + 0.1MKCl$. Scan rate = $0.01V\ s^{-1}$, reference electrode = Ag/AgCl.*

*Fig. 2. SEM images of the films electrodeposited in the nitrate solution for 1 hour; (a) -1.0 V top view;(b) -1V 45° tilted view; (c) -0.7V top view; (d) -0.7V 45° tilted view; (e)-0.6V top view; (f)-0.6V cross section, the two arrows mark the film thickness.*

*Fig. 3. XRD patterns of the ZnO films grown in the nitrate solution at different voltages: -1.0 V, -0.7 V and -0.6 V vs. Ag/AgCl.*

*Fig. 4. SEM images of the films electrodeposited in the peroxide solution for 1 hour; (a) -0.9V top view; (b) -0.9V 45° tilted view; (c) -0.5V top view. (d) -0.5V 45° tilted view; (e) -0.3V top view; (f) -0.3V cross sectional, the arrow marks the film thickness.*

*Fig. 5. XRD patterns of the ZnO films grown in the peroxide solution at different voltages: -0.9 V, -0.5 V and -0.3 V vs. Ag/AgCl.*

*Fig. 6. Gaussian deconvolution of the visible emission of electrodeposited ZnO films grown under different conditions.*

*Fig. 7. Relative weight of the different Gaussian components deconvoluted from the visible emission of the electrodeposited ZnO films grown under different conditions. Total area normalized to 1.*

*Fig. 8. Proposed transitions related to defect emission in the visible range* [32, 36].

*Fig. 9. Photoluminescence measured (in log scale) at room temperature of the sample grown at -0.5V in the peroxide solution, as grown and annealed at different temperatures.*

*Fig. 10. Relative weight for the Gaussian components deconvoluted from the emission of the electrodeposited ZnO film (grown at -0.5V in the peroxide solution). Total area normalized to 1*

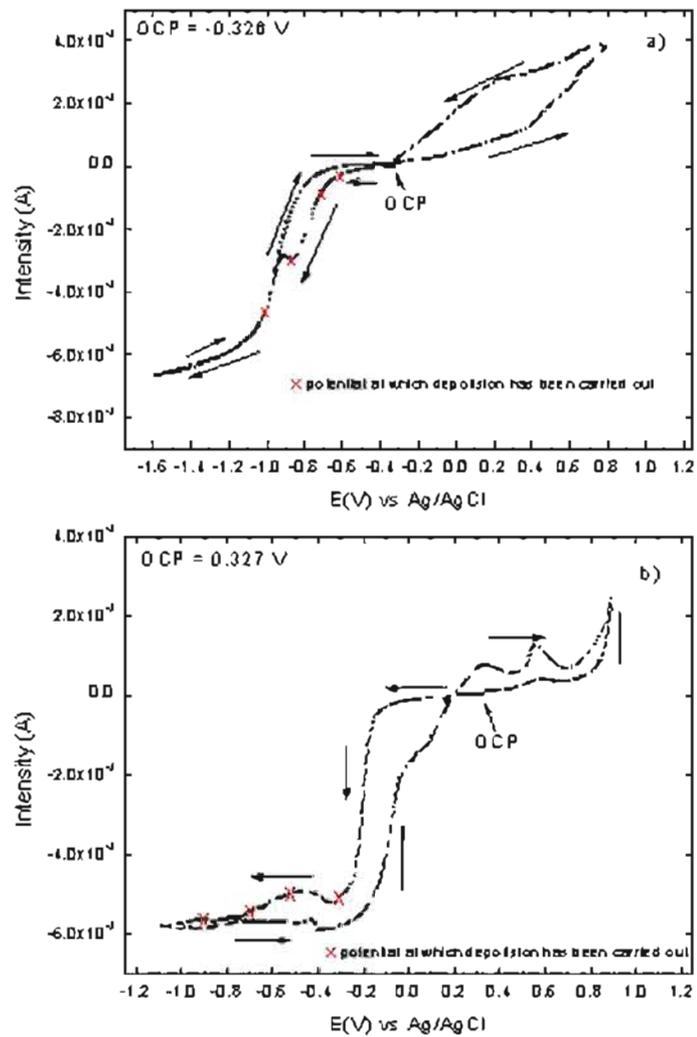

Fig. 1

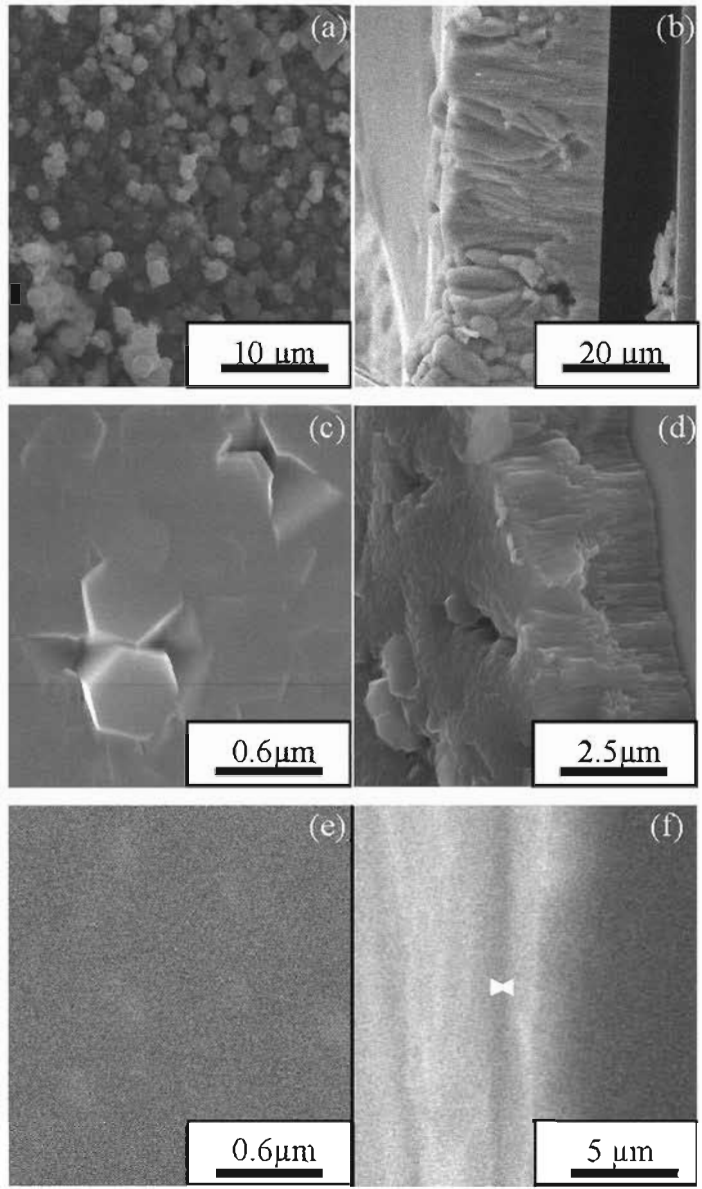

*Fig. 2.*

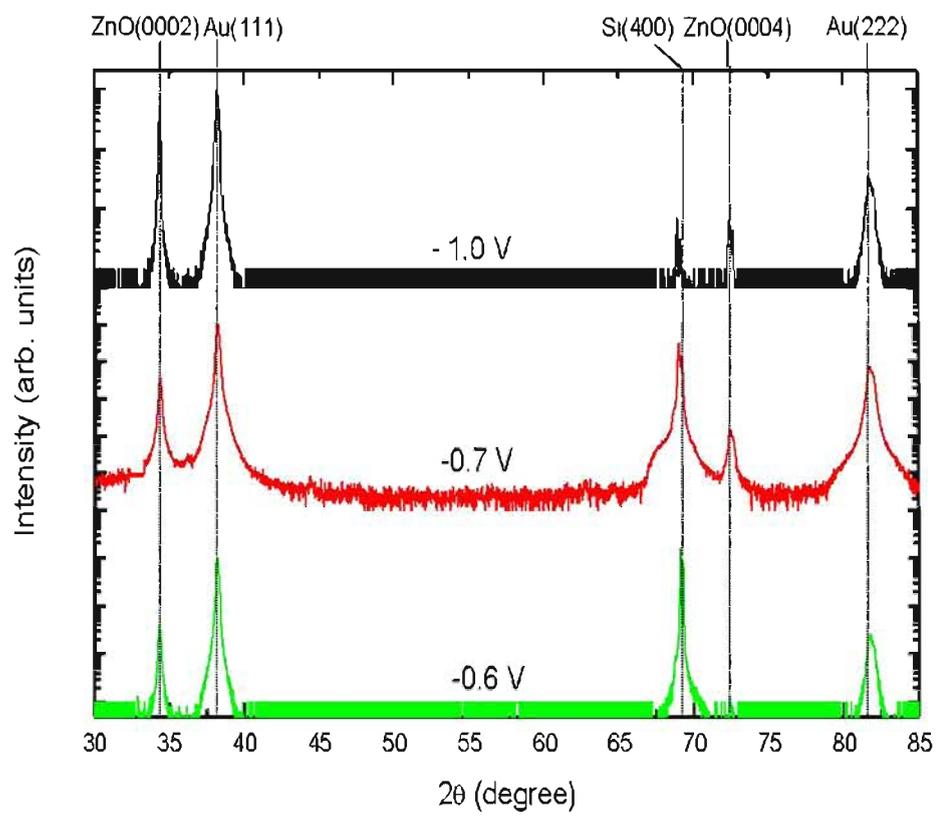

*Fig. 3.*

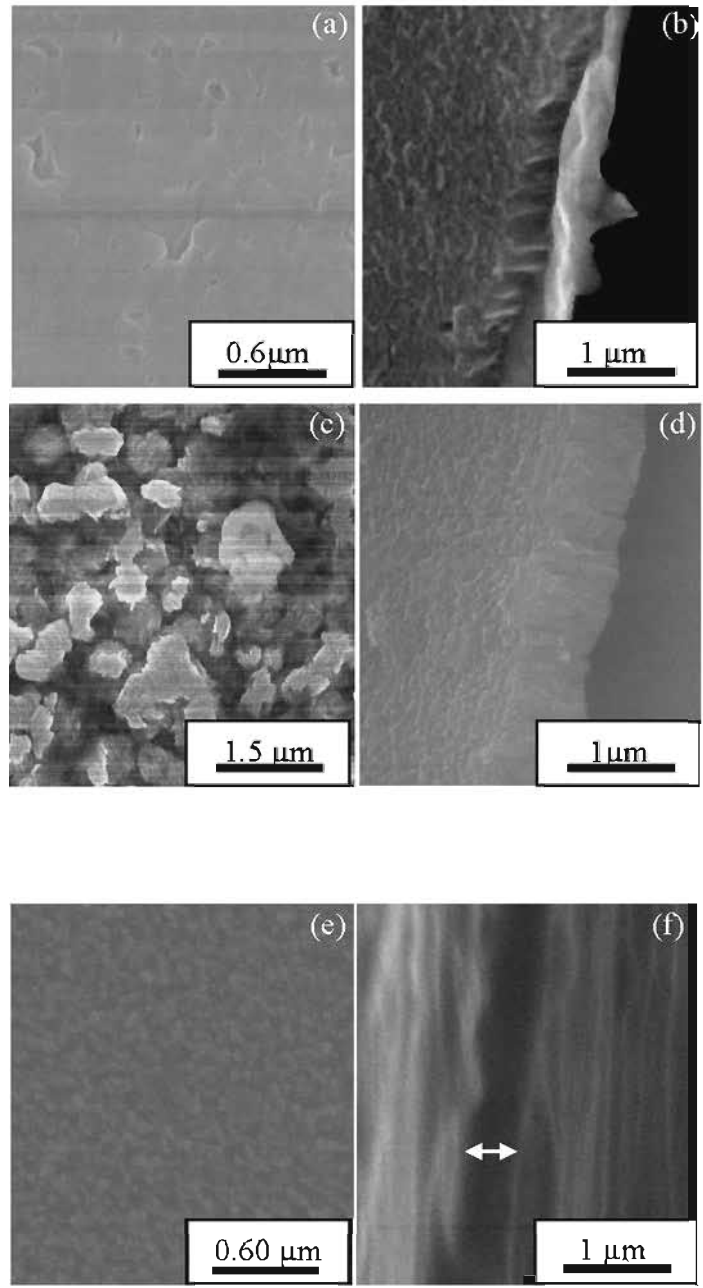

*Fig. 4.*

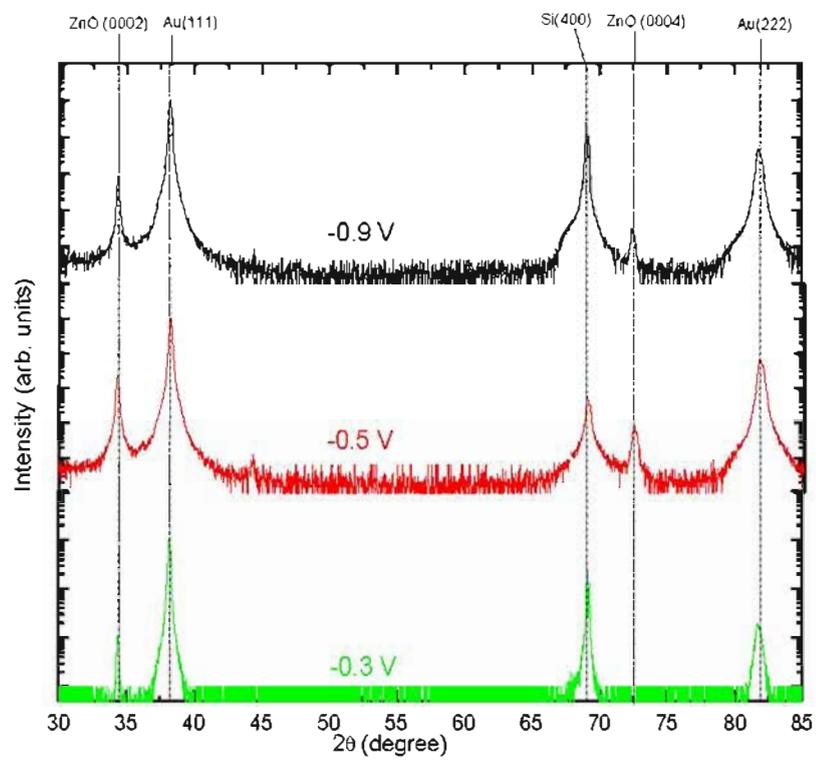

*Fig. 5.*

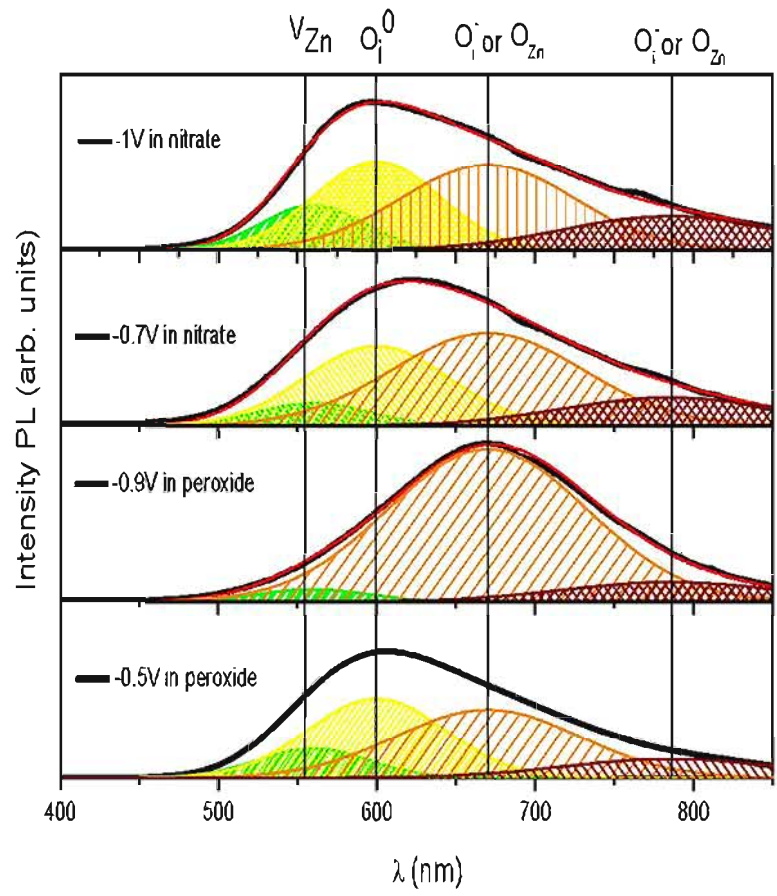

*Fig. 6.*

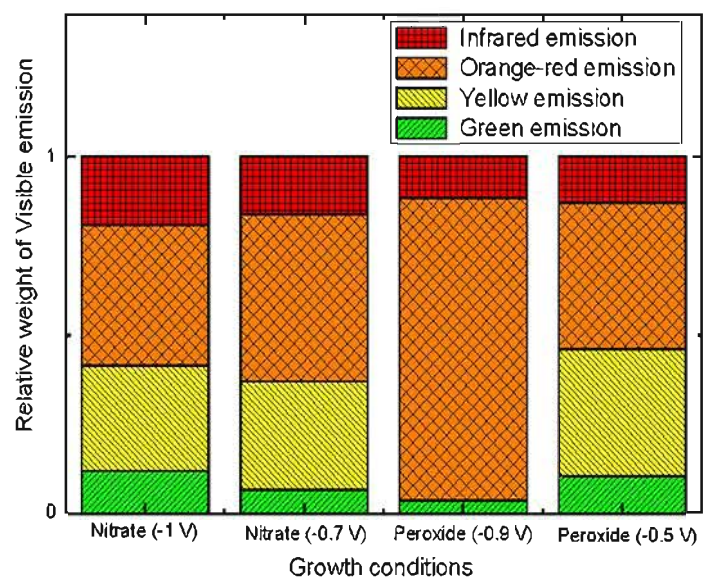

*Fig. 7.*

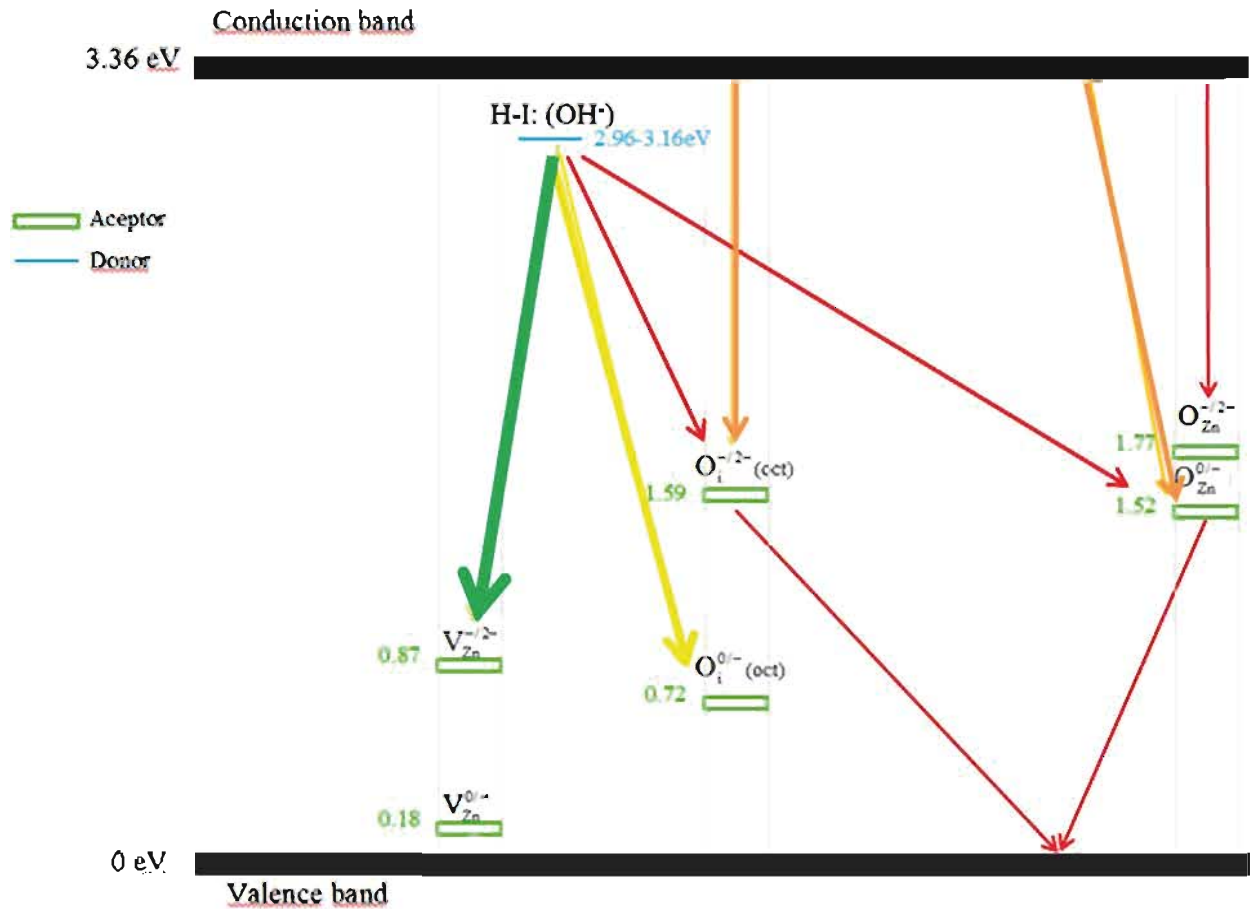

Fig. 8

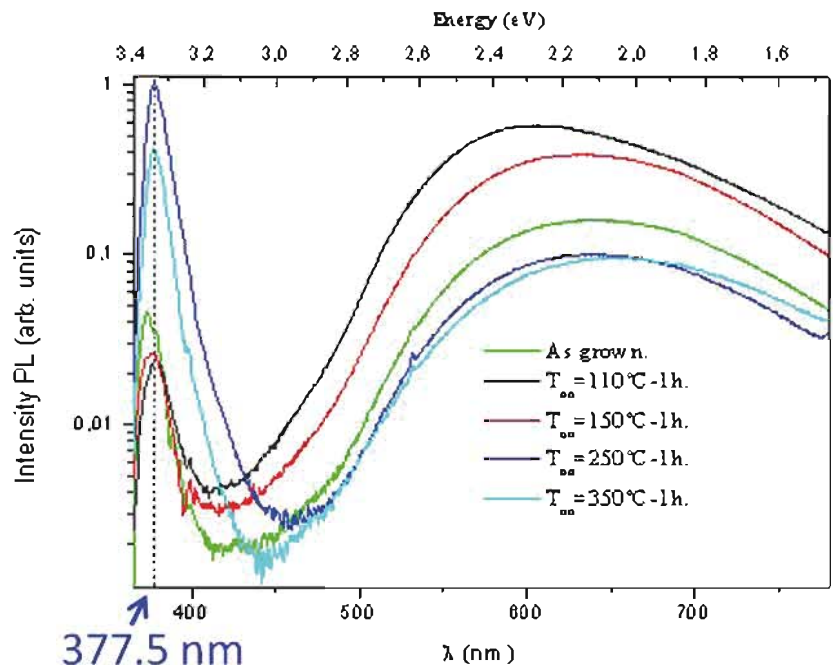

Fig. 9

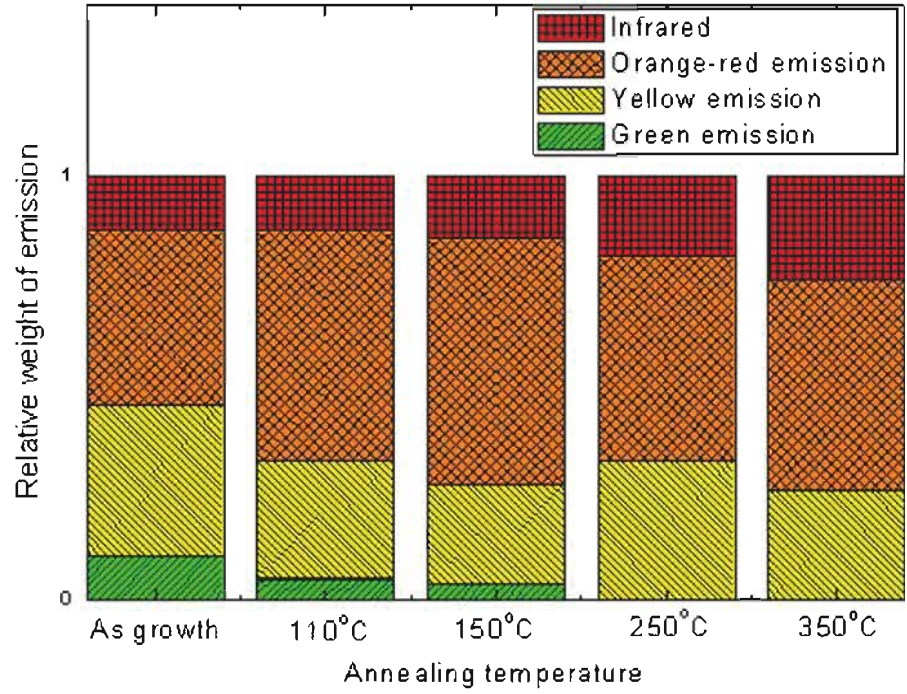

Fig. 10